\documentclass[11pt]{article}
\usepackage[T1]{fontenc}
\usepackage[utf8]{inputenc}
\usepackage[margin=1in]{geometry}
\usepackage{microtype}
\usepackage[hidelinks]{hyperref}
\usepackage{doi}
\usepackage{graphicx}
\usepackage{placeins}

\usepackage{amsmath,amssymb}
\usepackage{tikz-cd}
\usepackage{multicol}
\DeclareMathOperator{\sinc}{sinc}

\title{\textbf{Cherenkov Diffraction Radiation Interferometry for Beam Divergence Diagnostics}}
\author{A.\ Novokshonov\thanks{\texttt{artem.novokshonov@desy.de}}\\
Deutsches Elektronen-Synchrotron DESY, Germany}
\date{}

\begin{document}
\maketitle

\begin{abstract}
Cherenkov radiation (ChR) is widely used in charged-beam diagnostics, including bunch-length measurements and beam-loss monitoring. Synchrotron radiation (SR) interferometry is a well-established technique for measuring beam size. This work combines these approaches by proposing a ChR-based interferometric method for measuring beam divergence.
\end{abstract}

\section*{Introduction}
Cherenkov radiation is electromagnetic radiation emitted when a charged particle travels through a dielectric medium at a velocity greater than the phase velocity of light in that medium~\cite{cherenkov1937,frank1937,jelley1958}. It is widely used in accelerator physics, for example, in bunch-length monitors~\cite{davut2025}, beam-position monitors~\cite{clapp2024}, and beam-loss monitors~\cite{benitez2024}. Cherenkov diffraction radiation has also been directly observed in the visible range and investigated as a tool for noninvasive beam diagnostics~\cite{kieffer2018,alves2019}. A key property of ChR is that its emission angle depends on the particle velocity and the refractive index of the medium:
\begin{equation}
    \label{eq:theta_ch}
    \cos \theta_{\mathrm{ch}} = \frac{1}{n(\lambda) \, \beta},
\end{equation}
where $\theta_{\mathrm{ch}}$ is the emission angle, $n(\lambda)$ is the refractive index of the medium, and $\beta = v/c$ is the reduced particle velocity. Thus, for a bunch with a finite angular distribution observed at a fixed wavelength $\lambda_0$, the angular distribution of ChR is expected to reproduce the bunch angular distribution. For example, in the visible range, a lens could collect the light onto a camera sensor placed in its focal plane. The resulting image would be a convolution of the point-spread function (PSF) with the bunch angular distribution. However, modern accelerators typically have rms angular spreads of only a few microradians, which cannot be resolved by direct imaging because of the diffraction limit. To overcome this limitation, one can adopt a technique already used for beam-size measurements: interferometry.

Synchrotron radiation interferometry is a noninvasive method for measuring transverse beam size. In its usual form, synchrotron radiation from the beam passes through a double-slit interferometer, and the visibility of the resulting interference fringes is related to the transverse coherence of the light and, therefore, to the beam size~\cite{mitsuhashi1999,naito2006}. Because the method extracts beam-size information from fringe contrast rather than from direct imaging alone, it can provide higher sensitivity when the beam size becomes comparable to or smaller than the resolution limit of a conventional optical system.

\begin{figure}[!hbt]
    \centering
    \begin{minipage}[t]{0.48\textwidth}
        \centering
        \includegraphics[width=\textwidth]{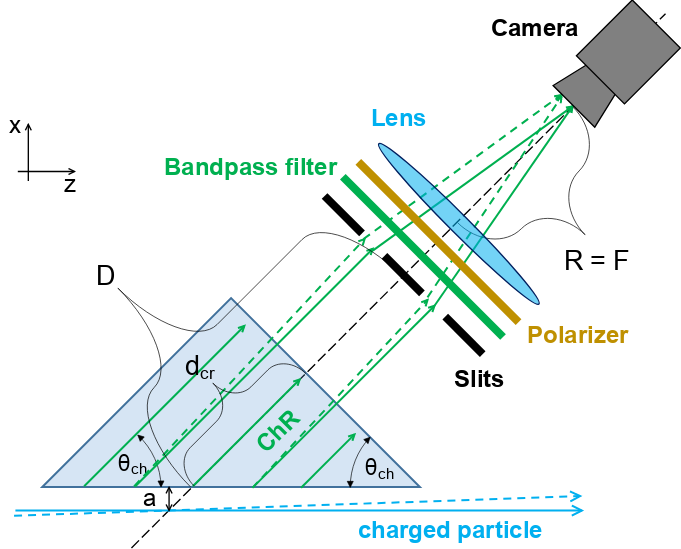}
        \caption{Schematic of the measurement.}
        \label{fig:scheme}
    \end{minipage}\hfill
    \begin{minipage}[t]{0.48\textwidth}
        \centering
        \includegraphics[width=\textwidth]{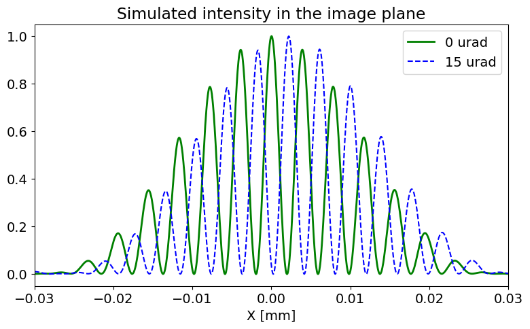}
        \caption{Simulated interferograms from two particles with different trajectories.}
        \label{fig:PSFs}
    \end{minipage}
\end{figure}

\begin{figure}[!ht]
    \centering
    \includegraphics[width=0.8\textwidth]{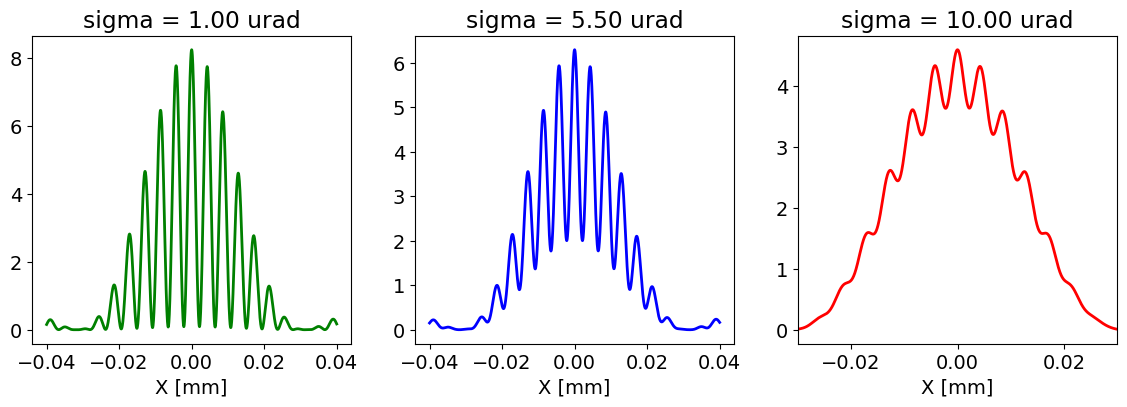}
    \caption{Examples of interferograms from bunches.}
    \label{fig:examples_beams}
\end{figure}

We propose using ChR instead of SR in this interferometric technique. A schematic of the setup is shown in Fig.~\ref{fig:scheme}. A charged particle passes near a crystal, parallel to its edge, as indicated by the solid blue arrow. The distance between the particle trajectory and the crystal is denoted by $a$. Cherenkov radiation is then generated inside the crystal. The observation wavelength, selected by the bandpass filter, is $\lambda_0$; this wavelength determines the ChR emission angle $\theta_{\mathrm{ch}}$ and, consequently, the geometry of the crystal. After leaving the crystal, the light passes through the slits, bandpass filter, polarizer, and lens before being focused onto the camera sensor in the focal plane. The slit separation is denoted by $d$, and the slit half-width by $w_{\mathrm{sl}}$. It is important to note that $D \gg d_{\mathrm{cr}}$ and that only the far-field approximation is considered.
Then a normalized interferogram recorded by the camera is shown as the solid green curve in Fig.~\ref{fig:PSFs}; the horizontal axis corresponds to the camera sensor plane. Now consider another particle whose trajectory makes a small angle, for example $\alpha = 15\,\mu\mathrm{rad}$, with respect to the initial trajectory. The corresponding interferogram is shifted, as illustrated by the blue dashed line in Fig.~\ref{fig:PSFs}. For a bunch of particles with a Gaussian angular distribution, the recorded image should therefore be given by the convolution of that distribution with the single-particle interferogram. The resulting interferograms are shown in Fig.~\ref{fig:examples_beams}; the larger the beam divergence, the more smeared out the interferogram becomes.

The visibility of the interferogram is defined as
\begin{equation}
    \label{eq:visibility}
    V = \frac{I_{\max} - I_{\min}}{I_{\max} + I_{\min}},
\end{equation}
where $I_{\max}$ and $I_{\min}$ are the intensities at a maximum and a minimum, respectively. The visibility therefore contains information about the beam angular distribution.

It should be noted that an additional lens or objective is typically used to provide magnification. As shown in Figs.~\ref{fig:examples_beams} and~\ref{fig:PSFs}, the full interferogram occupies only about $40\,\mu\mathrm{m}$. Commercial cameras typically have pixel sizes no smaller than about $2\,\mu\mathrm{m}$. At this resolution, fitting the interferogram would be impossible. Therefore, a second lens is required for magnification. In the simulations below, a magnification of $10\times$ is used. The role of the lenses is discussed further below in the context of monitor resolution.

\section*{Theory}
We first consider the well-known part of the setup: the interferometer itself, which consists of the slits, lens, and camera. In principle, the polarizer can be omitted because ChR is radially polarized. For greater precision, however, it can be included because the slits have finite height and may therefore cut off part of the ChR cone. This part of the setup is well established~\cite{mitsuhashi1999,naito2006,torino2014} and is not discussed here in detail. For rectangular slits, the interference pattern is described by
\begin{equation}
    \label{eq:int_fit}
    I = I_0 \, \operatorname{sinc}\!\left( \frac{2 \pi w_{\mathrm{sl}}}{\lambda_0 R} x \right)
        \left[ 1 + V \cos \left( \frac{2 \pi d}{\lambda_0 R} x + \phi \right) \right],
\end{equation}
where $I_0$ is the sum of the intensities from both slits and $V$ is the visibility. The visibility is related to the complex degree of spatial coherence $\gamma$ by
\begin{equation}
    \label{eq:vis_gamma_1}
    V = \frac{2 \sqrt{I_1 I_2}}{I_1 + I_2} \, |\gamma|,
\end{equation}
where $I_1$ and $I_2$ are the intensities transmitted through the two slits. For simplicity, we consider only the case $I_1 = I_2$, for which $V = |\gamma|$. According to the van Cittert--Zernike theorem~\cite{vancittert1934,bornwolf1980}, the complex degree of spatial coherence is the Fourier transform of the source distribution, which in our case is the bunch angular distribution. This is the main difference between SR and ChR interferometry: in conventional SR interferometry, $\gamma$ depends on the transverse position distribution of the radiation source, whereas in the present case it should depend on the angular distribution.

\begin{figure}[!hbt]
    \centering
    \includegraphics[width=0.6\textwidth]{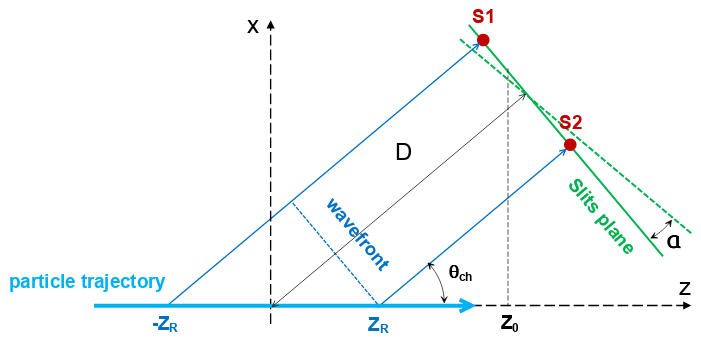}
    \caption{Geometry used for the phase-difference calculation.}
    \label{fig:phase_diff}
\end{figure}

Following the logic of the van Cittert--Zernike theorem, we need to find the cross-correlation between the fields at the slits $S_1$ and $S_2$ (see Fig.~\ref{fig:phase_diff}).
To derive $\gamma$, let us first represent the Cherenkov wavefront as a sum of plane waves, with the radiation source located at the origin of the coordinate system. A single electromagnetic plane wave emitted at an angle $\theta_{\mathrm{ch}}$ can be written as
\begin{equation}
    \label{eq:single_plane_wave}
    \mathbf{H}(\mathbf{r}, \omega) = A_0 \, e^{i\mathbf{k} \cdot \mathbf{r}},
\end{equation}
where $\mathbf{k} = \frac{n \omega}{c} \{\sin\theta_{\mathrm{ch}}\cos\phi, \, \sin\theta_{\mathrm{ch}}\sin\phi, \, \cos\theta_{\mathrm{ch}} \}$ and $\phi$ is the azimuthal angle. The full ChR cone is then described by
\begin{equation}
    \label{eq:chr_cone1}
    \mathbf{H}(x,y,z,\omega) = \int_0^{2\pi} A_0 \, e^{i k (x\sin\theta_{\mathrm{ch}}\cos\phi + y\sin\theta_{\mathrm{ch}}\sin\phi + z\cos\theta_{\mathrm{ch}})} \, \mathrm{d}\phi.
\end{equation}
Substituting $x = r \cos\alpha$ and $y = r\sin\alpha$, we obtain
\begin{equation}
    \label{eq:chr_cone2}
    \mathbf{H}(r,z,\alpha,\omega) = A_0 \, e^{ikz\cos\theta_{\mathrm{ch}}} \int_0^{2\pi} e^{ikr\sin\theta_{\mathrm{ch}}\cos(\phi - \alpha)} \, \mathrm{d}\phi.
\end{equation}
Evaluating the integral gives
\begin{equation}
    \label{eq:chr_cone3}
    \mathbf{H}(r,z,\omega) = A(r) \, e^{ikz \cos\theta_{\mathrm{ch}}},
\end{equation}
where $A(r) = 2 \pi A_0 \, J_0(k r \sin\theta_{\mathrm{ch}})$. Now assume that there are two radiation sources on the $Z$ axis with coordinates $\pm Z_R$, and that the slit plane is slightly tilted, as shown by the dashed green line in Fig.~\ref{fig:scheme}. This slit-plane tilt effectively simulates a tilt of the particle trajectory, but the geometry is easier to depict in this form. The radiation intensity also depends on $\alpha$ because the bunch is ultimately assumed to have an angular distribution in $\alpha$. Two important assumptions are used here: $\alpha \to 0$ and $D \gg Z_R$. The fields at the slits are then
\begin{equation}
    \begin{gathered}
        \label{eq:fields_in_slits_1}
            \mathbf{H}(S_1) = A(r_{S_1}) e^{ik(Z_0 + Z_R - \Delta Z(\alpha)) \cos\theta_{\mathrm{ch}}} \\
            \mathbf{H}(S_2) = A(r_{S_2}) e^{ik(Z_0 - Z_R + \Delta Z(\alpha)) \cos\theta_{\mathrm{ch}}}
    \end{gathered}
\end{equation}
We now calculate the mutual coherence function:
\begin{equation}
    \label{eq:mut_coh_func}
    \langle H(S_1) H^*(S_2) \rangle = \langle A(r_{S_1}) \, A^*(r_{S_2}) \rangle \, e^{2ik(Z_R - \Delta Z(\alpha)) \cos\theta_{\mathrm{ch}}}.
\end{equation}
Here, $\langle A(r_{S_1}) \, A^*(r_{S_2}) \rangle$ is essentially the source intensity, which ultimately depends on $\alpha$, i.e., $I(\alpha)$. The normalized complex degree of spatial coherence is therefore
\begin{equation}
    \label{eq:gamma_1}
        \gamma = \frac{\int I(\alpha) \, e^{2ik (Z_R - \Delta Z(\alpha)) \cos\theta_{\mathrm{ch}}} \, \mathrm{d}\alpha}{\int I(\alpha) \, \mathrm{d} \alpha}.
\end{equation}
As mentioned above, $d$ is the slit separation, so $|Z_R| = d / (2 \, \sin\theta_{\mathrm{ch}})$. On the slit side, the corresponding displacement is $\Delta Z(\alpha)$. Since $\alpha \to 0$, we may write $\Delta Z(\alpha) = d \, \alpha / 2$. We can therefore rewrite the integral as
\begin{equation}
    \label{eq:gamma_2}
    \gamma = e^{ik d \tan\theta_{\mathrm{ch}}} \frac{\int I(\alpha) \, e^{- ik d \, \alpha \cos\theta_{\mathrm{ch}}} \mathrm{d}\alpha}{\int I(\alpha) \, \mathrm{d}\alpha}.
\end{equation}
We now assume that the intensity distribution is Gaussian, with standard deviation $\sigma_{\alpha}$. Solving the integral and simplifying gives
\begin{equation}
    \label{eq:gamma_abs}
    |\gamma| = e^{- (k \, d \, \cos\theta_{\mathrm{ch}} \, \sigma_{\alpha})^2 / 2},
\end{equation}
and therefore
\begin{equation}
    \label{eq:sig_gamma}
    \sigma_{\alpha} = \frac{\sqrt{2 \log(|\gamma|^{-1})}}{k \, d \cos\theta_{\mathrm{ch}}}.
\end{equation}
Thus, the interferogram can be fitted with Eq.~\ref{eq:int_fit}. After accounting for the relative intensities from the two slits, the measured visibility can then be converted into the beam divergence.

\section*{Simulation}
\begin{figure}[!ht]
    \centering
    \includegraphics[width=0.8\textwidth]{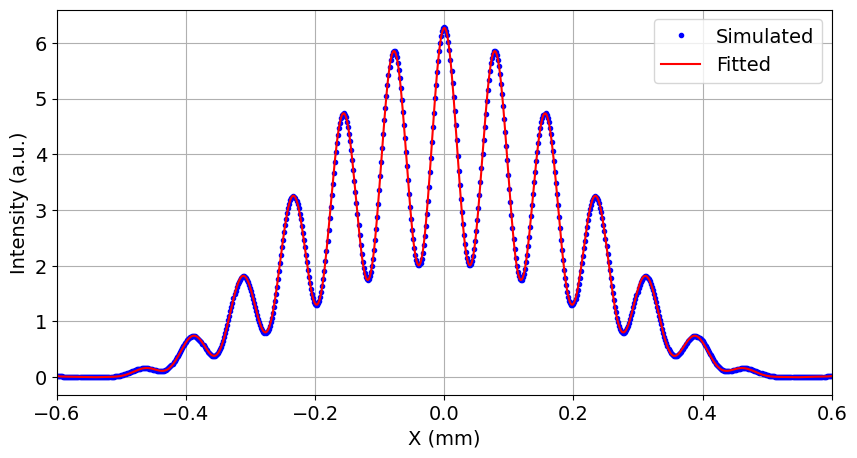}
    \caption{Simulated and fitted interferogram.}
    \label{fig:sim_fit_int}
\end{figure}

\begin{figure}[!ht]
    \centering
    \begin{minipage}[t]{0.32\textwidth}
        \centering
        \includegraphics[width=\textwidth]{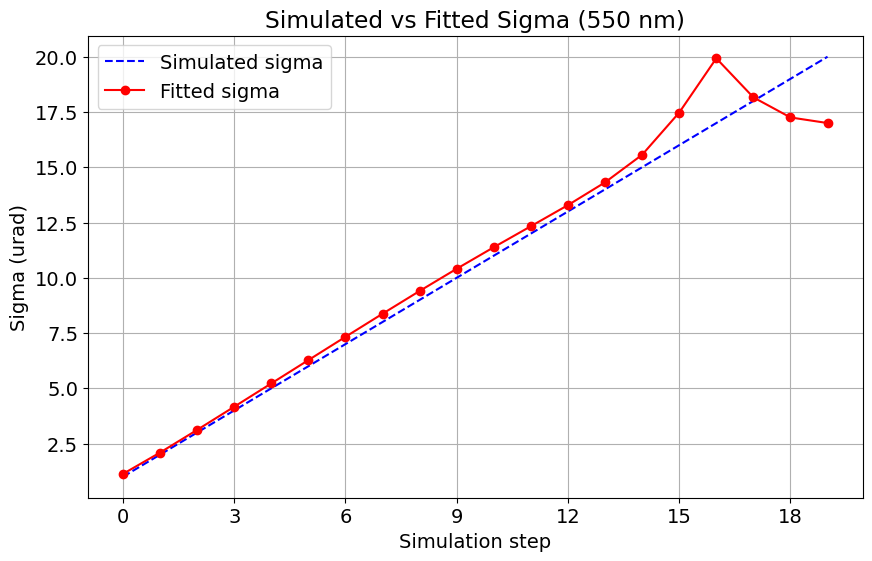}
        \textbf{(a)}
    \end{minipage}\hfill
    \begin{minipage}[t]{0.32\textwidth}
        \centering
        \includegraphics[width=\textwidth]{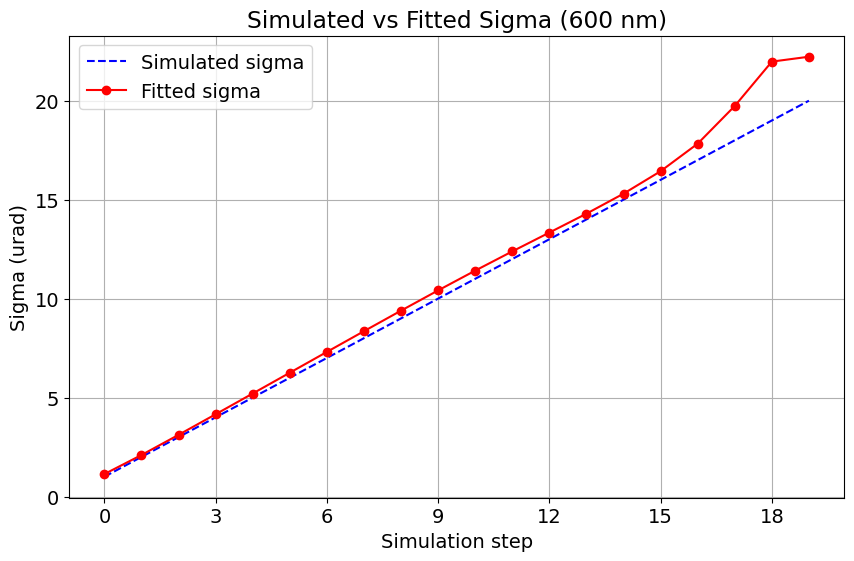}
        \textbf{(b)}
    \end{minipage}\hfill
    \begin{minipage}[t]{0.32\textwidth}
        \centering
        \includegraphics[width=\textwidth]{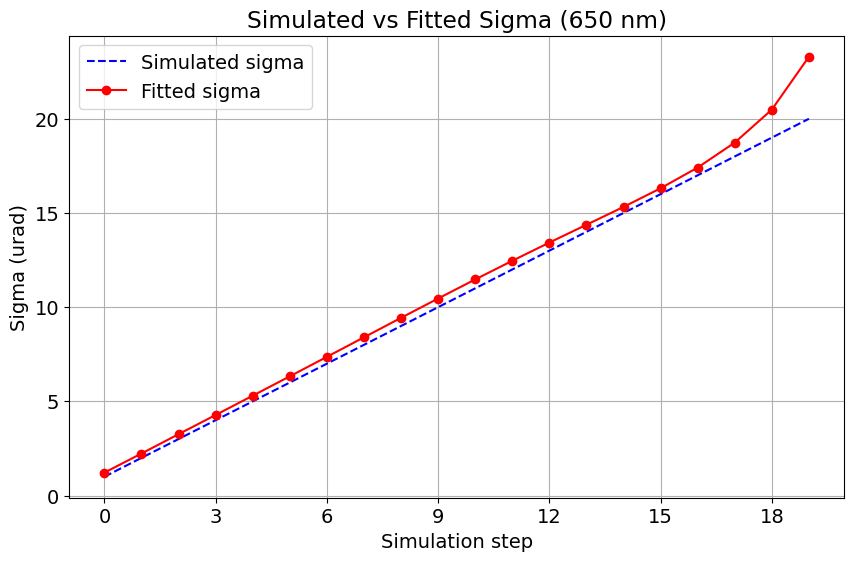}
        \textbf{(c)}
    \end{minipage}
    \caption{Bunch divergences reconstructed from simulated interferograms.}
    \label{fig:restored_divs}
\end{figure}

To simulate the measurement process, we use the polarization-current model described by Karlovets and Potylitsyn~\cite{karlovets2009}. In particular, we use the following expression:
\begin{equation}
    \label{eq:pol_cur}
    \begin{gathered}
        \mathbf{H}^{R}(\mathbf{r}, \omega)
        = \frac{e \omega \sqrt{\varepsilon}(\varepsilon - 1)a}{2 \pi v c \gamma^2} \times
        \frac{e^{i r \sqrt{\varepsilon}\omega/c}}{r} \times
        \frac{e^{-i d_1 \frac{\omega}{c}(\beta^{-1} - \sqrt{\varepsilon}\cos\Theta)}
        - e^{i d_2 \frac{\omega}{c}(\beta^{-1} - \sqrt{\varepsilon}\cos\Theta)}}
        {(\beta^{-1} - \sqrt{\varepsilon}\cos\Theta)
        \left[1 - \beta^2 + (\beta \sqrt{\varepsilon}\sin\Theta)^2\right]} \\
        \quad \times \Bigg[
        \sin\Theta\left(\gamma^{-1} - \beta\gamma\sqrt{\varepsilon}\cos\Theta\right)
        J_0\!\left(\frac{a\omega}{c}\sqrt{\varepsilon}\sin\Theta\right)
        K_1\!\left(\frac{a\omega}{v\gamma}\right) \\
        \quad - \left(\cos\Theta + \beta\sqrt{\varepsilon}\sin^2\Theta\right)
        J_1\!\left(\frac{a\omega}{c}\sqrt{\varepsilon}\sin\Theta\right)
        K_0\!\left(\frac{a\omega}{v\gamma}\right)
        \Bigg] \\
        \quad \times \{\sin\phi, -\cos\phi, 0\}.
    \end{gathered}
\end{equation}
All simulations are performed for electrons. The initial wavefront from a single electron is simulated in the plane of the crystal--air boundary and then propagated through the boundary, lens, and slits to the sensor plane. A single-particle interferogram has already been shown in Fig.~\ref{fig:PSFs}. The electron energy used in the simulations is $130\,\mathrm{MeV}$. An example of a simulated and fitted interferogram from an electron bunch is shown in Fig.~\ref{fig:sim_fit_int}.
The main simulation parameters are listed in Table~\ref{tab:sim_params}.
\begin{table}[!ht]
    \centering
    \caption{Simulation parameters.}
    \label{tab:sim_params}
    \begin{tabular}{c|cccccc}
        \hline
        Parameter & $\lambda$ & $D$ & $F$ & $d$ & $w_{\mathrm{sl}}$ & $\sigma_{\alpha}$\\
        \hline
        Value & $550\,\mathrm{nm}$ & $10\,\mathrm{m}$ & $0.2\,\mathrm{m}$ & $14\,\mathrm{mm}$ & $2\,\mathrm{mm}$ & $5\,\mu\mathrm{rad}$\\
        \hline
    \end{tabular}
\end{table}
Fused silica is chosen as the crystal medium, and the Sellmeier equation is used for the refractive-index calculations~\cite{malitson1965}. We now simulate several bunches and reconstruct their divergences. Figure~\ref{fig:restored_divs} shows three simulation results obtained using the parameters $D$, $F$, $d$, and $w_{\mathrm{sl}}$ listed in Table~\ref{tab:sim_params}, but with different wavelengths: (a) $550\,\mathrm{nm}$, (b) $600\,\mathrm{nm}$, and (c) $650\,\mathrm{nm}$. The blue dashed line in all three plots represents the simulated beam divergence, which is the same in each case: $1$--$20\,\mu\mathrm{rad}$. These results indicate that longer wavelengths allow larger beam divergences to be measured, as expected. For sufficiently large divergences, however, the visibility approaches zero, which sets the measurement limit. The slit width and separation can be adjusted to optimize the sensitivity for a given divergence range, but we do not discuss this optimization here because it is a well-known feature of interferometric measurements. Another noticeable feature in all three plots is a small discrepancy between the fit and the simulation roughly in the middle of the considered range, where the fitted values are slightly larger. This discrepancy does not yet have a clear explanation. However, because the largest difference is below $\approx 5\%$, it can be neglected for now.

\begin{figure}[ht]
    \centering
    \includegraphics[width=0.55\textwidth]{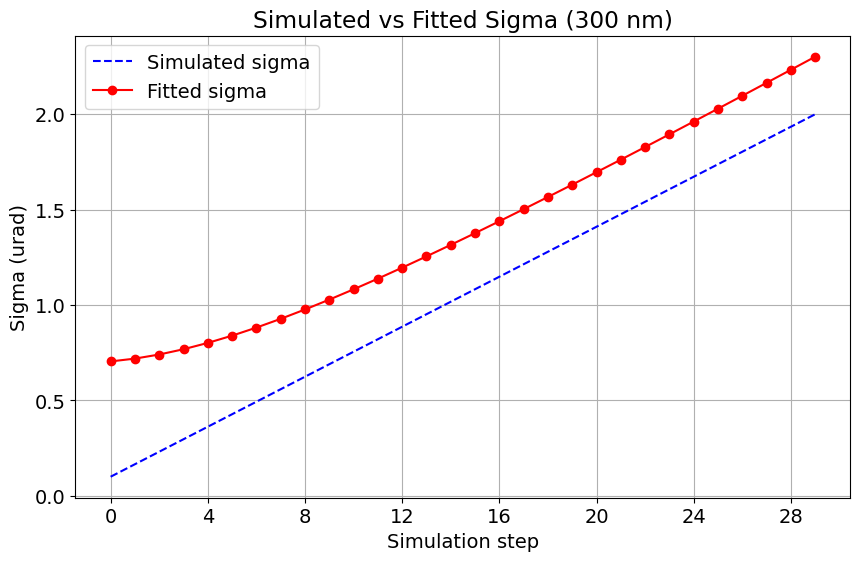}
    \caption{Estimated resolution.}
    \label{fig:resolution}
\end{figure}

A further question is the smallest divergence that can be measured. This brings us back to the lenses and objectives, because they introduce a point-spread function (PSF) that limits the resolution. The observed interferogram is therefore a convolution of the ideal interference pattern with the PSF of the optical system, which can be approximated by an Airy disk. A precise PSF depends on the specific optics used; here, we use a simple estimate. Commercial objectives typically have PSF diameters on the order of a few micrometers. For two optical elements, the effective PSF diameter can be estimated as $r_{\mathrm{tot}} = \sqrt{r_1^2 + r_2^2}$. Assuming a PSF of about $3\,\mu\mathrm{m}$ for each objective gives an effective value of about $4.5\,\mu\mathrm{m}$. To access smaller divergences, a shorter wavelength is required; here, we use $300\,\mathrm{nm}$, since substantially shorter wavelengths would make the optical setup more challenging. The interferograms are then simulated, convolved with the PSF, and fitted for divergences in the range $0.1$--$2\,\mu\mathrm{rad}$. The simulation result is shown in Fig.~\ref{fig:resolution}. Below about $1\,\mu\mathrm{rad}$, the fitted divergences begin to deviate from the expected linear dependence. Going to even shorter wavelengths could further improve this limit, but it may also introduce additional challenges for the optical setup. Another possible improvement is to introduce an intensity imbalance between the two slits, a technique known from SR interferometry~\cite{boland2012}. However, this technique is not expected to improve the resolution substantially and has already been discussed in the SR-interferometry literature, so it is not considered further here. We therefore conclude that the resolution of the proposed monitor, in the optical range considered here, lies in the sub-microradian range.

\section*{Intensity}
\begin{figure}[ht]
    \centering
    \includegraphics[width=0.9\textwidth]{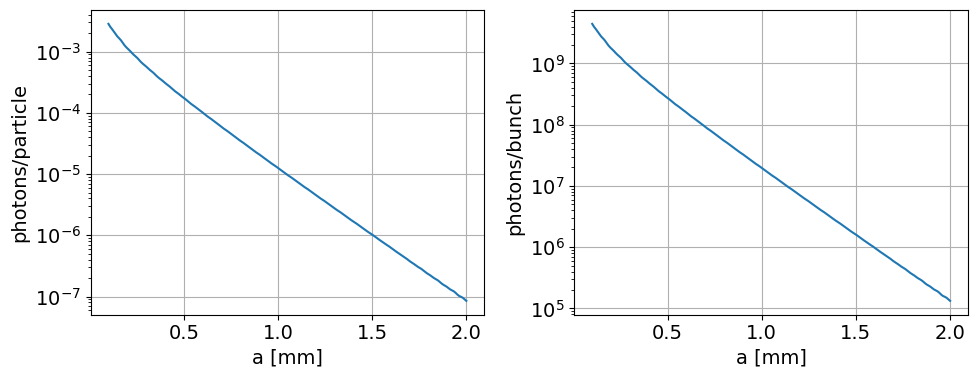}
    \caption{Intensity estimation: left - per one electron; right - per bunch.}
    \label{fig:intensity_est}
\end{figure}

An important question for diffraction radiation is its intensity, or, in other words, whether a sufficient number of photons is produced. This intensity depends on the particle--radiator distance $a$ (see Fig.~\ref{fig:scheme}). To estimate it, let us start from the angular spectral distribution:
\begin{equation}
    \label{eq:ang_spect_dist}
    \frac{d^2 W}{d\omega \, d\Omega} = c r^2 \sqrt{\varepsilon} \left| \mathbf{H}^{R} \right|^2.
\end{equation}
In our case, the angular acceptance is defined by the slits aperture. We can therefore integrate over $d\Omega$:
\begin{equation}
    \label{eq:spect_dist1}
    \frac{d W}{d\omega} = c r^2 \sqrt{\varepsilon} \int \left| \mathbf{H}^{R} \right|^2 \, d\Omega =
    c r^2 \sqrt{\varepsilon} \iint \left| \mathbf{H}^{R} \right|^2 \sin\theta \, d\theta \, d\phi.
\end{equation}
We now return to Eq.~\ref{eq:pol_cur} and find its squared modulus:
\begin{equation}
    \label{eq:pol_cur_abs}
    \begin{gathered}
        |\mathbf{H}^R|^2 =
        \left(\frac{e \omega a}{2 \pi v c \gamma^2}\right)^2
        \frac{\varepsilon (\varepsilon - 1)^2}{r^2}
        \left(\frac{d_t \, \omega}{c}\right)^2
        \frac{\sinc^2\left( \frac{d_t \, \omega}{2 c} (\beta^{-1} - \sqrt{\varepsilon} \cos\theta) \right)}
             {|1 - \beta^2 +\beta^2 \varepsilon \sin^2\theta|^2} \\
        \times
        \Biggl[
            \sin\theta \left( \gamma^{-1} -\beta\gamma\sqrt{\varepsilon} \cos\theta \right)
            J_0 \left( a \frac{\omega}{c} \sqrt{\varepsilon} \sin\theta \right)
            k_1 \left( a \frac{\omega}{v \gamma} \right) \\
            -
            \left( \cos\theta + \beta\sqrt{\varepsilon} \sin^2\theta \right)
            J_1 \left( a \frac{\omega}{c} \sqrt{\varepsilon} \sin\theta \right)
            k_0 \left( a \frac{\omega}{v \gamma} \right)
        \Biggr]^2,
    \end{gathered}
\end{equation}
where $d_t = d_1 + d_2$ is the total width of radiator. One can see there is no dependence on $\phi$ therefore by substituting it into Eq.~\ref{eq:spect_dist1} and integrating over $\phi$:
\begin{equation}
    \label{eq:spect_dist2}
    \frac{d W}{d\omega} = \Delta\phi_s \, c r^2 \sqrt{\varepsilon} \int \left| \mathbf{H}^{R} \right|^2 \sin\theta \, d\theta,
\end{equation}
with $\Delta\phi_s = \phi_1 - \phi_0$ is the azimuthal aperture of the slit. Let us consider now large width of radiator, the interference term becomes:
\begin{equation}
    \left(\frac{d_t \, \omega}{c}\right)^2
    \sinc^2\left(\frac{d_t \, \omega}{2 c} (\beta^{-1} - \sqrt{\varepsilon}\cos\theta)\right)
    \to
    2 \pi \frac{d_t \, \omega}{c} \delta\left(\beta^{-1} - \sqrt{\varepsilon}\cos\theta \right).
\end{equation}
The whole integral is now dependent on the $\delta$-function, which is equal to one only at the Cherenkov angle. Hence we should only evaluate it at the angle. So the final spectral distribution is:
\begin{equation}
    \label{eq:spect_int1}
    \begin{gathered}
    \frac{dW}{d\omega} = \Delta\phi_s \,
    \left(\frac{e \omega a}{c^2 \gamma^2}\right)^2
    \frac{\varepsilon}{\beta^4}
    \left(\frac{d_t \, \omega}{2 \pi}\right) \times
    \Biggl[
        \gamma\sin\theta_{ch}
        J_0 \left( a \frac{\omega}{c} \sqrt{\varepsilon} \sin\theta_{ch} \right)
        K_1 \left( a \frac{\omega}{v\gamma} \right) \\
        + \,
        \beta^{-1} \sqrt{\varepsilon}
        J_1 \left( a \frac{\omega}{c} \sqrt{\varepsilon} \sin\theta_{ch} \right)
        K_0 \left( a \frac{\omega}{v\gamma} \right)
    \Biggr]^2.
    \end{gathered}
\end{equation}
And as the last step lets switch to $\lambda$ and SI units as the initial formula is in Gaussian SGC units:
\begin{equation}
    \label{eq:spec_int2}
    \begin{gathered}
    \frac{dW}{d\lambda} = \Delta\phi_s
    \frac{2 \pi^2 e^2 a^2 d_t \varepsilon}{\varepsilon_0 \gamma^4 \beta^2 \lambda^5} \times
    \Biggl[
        \gamma\sin\theta_{ch}
        J_0 \left(\frac{2 \pi a}{\lambda} \sqrt{\varepsilon} \sin\theta_{ch} \right)
        K_1 \left(\frac{2 \pi a}{\beta\gamma\lambda} \right) \\
        + \,
        \beta^{-1} \sqrt{\varepsilon}
        J_1 \left(\frac{2 \pi a}{\lambda} \sqrt{\varepsilon} \sin\theta_{ch} \right)
        K_0 \left(\frac{2 \pi a}{\beta\gamma\lambda} \right)
    \Biggr]^2
    \end{gathered}
\end{equation}
With this formula, we can calculate the number of photons. For example, consider an electron beam with an energy of $2.4\,\mathrm{GeV}$ and a charge of $250\,\mathrm{pC}$, which are typical parameters of the European XFEL beam after its second linac section. Suppose that we want to estimate the number of photons emitted within the wavelength range $\lambda = 550 \pm 10\,\mathrm{nm}$. The radiator is $40\,\mathrm{mm}$ thick, the distance to the slit plane is $4\,\mathrm{m}$, and the slit width and height are $2\,\mathrm{mm}$ and $10\,\mathrm{mm}$, respectively. The impact parameter $a$ is then varied over the range $0.1$--$2\,\mathrm{mm}$. The results are shown in Fig.~\ref{fig:intensity_est}. The left panel shows the number of photons per electron, whereas the right panel shows the number per bunch; both values are calculated for a single slit. The photon yield should be sufficient at a distance of around $1\,\mathrm{mm}$. For comparison, Kieffer \emph{et al.}~\cite{kieffer2018} reported $0.8\,\mathrm{mm}$ as an optimal impact parameter. In that experiment, however, the impact parameter was mainly constrained by machine protection: the measurements were performed at the Cornell Electron Storage Ring (CESR), and because the machine is circular, moving the radiator closer to the beam would lead to beam loss. At the European XFEL, this should be less problematic; moreover, the signal could be integrated over several bunches in a bunch train to increase the intensity.

\begin{figure}[!ht]
    \centering
    \includegraphics[width=0.9\textwidth]{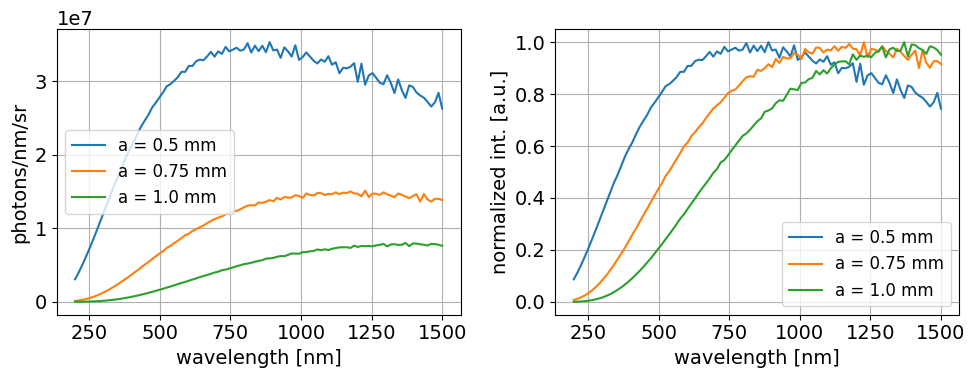}
    \caption{Dependence of the spectrum on impact parameter: left, spectral angular distributions; right, spectral angular distributions normalized to their maxima.}
    \label{fig:specs_est}
\end{figure}

Another important question is the radiation spectrum, because it also depends on the impact parameter. Let us consider three different impact parameters: $0.5$, $0.75$, and $1\,\mathrm{mm}$. All other parameters, such as the beam energy and charge, are kept the same as before. The resulting spectral angular distributions are shown in Fig.~\ref{fig:specs_est}. The distributions are calculated per electron: the left panel shows the spectral angular distributions, while the right panel shows the same distributions normalized to their maxima. As expected, the intensity increases as the impact parameter decreases. In addition, the spectrum shifts toward longer wavelengths for larger impact-parameter values. Therefore, the final choice of impact parameter should also account for the detector sensitivity. Simply choosing the smallest possible impact parameter would increase the photon yield, but those photons might not be detected efficiently if the detector quantum efficiency is low in the corresponding wavelength range. This behavior has also been discussed, for example, in Ref.~\cite{alves2019}.

\section*{Conclusion}
This work proposes a new noninvasive technique for beam-divergence measurements based on Cherenkov radiation. The same type of radiation can also be used for beam-size measurements~\cite{novokshonov2025} and can be generated noninvasively~\cite{alves2019}. Together, these properties suggest that Cherenkov radiation could be used to measure beam emittance noninvasively. Because the resolutions of these monitors can reach the microradian and micrometer ranges, respectively, they may be suitable for modern accelerator facilities, such as the European XFEL. The main concern for the proposed monitor is its sensitivity, but the estimates presented in this work indicate that the photon yield should be sufficient. In addition, the signal level can be adjusted either by reducing the impact parameter, i.e., moving the beam closer to the radiator, or by increasing the number of bunches used for integration. Both approaches depend strongly on the acceptable beam losses downstream of the monitor.



\begin{thebibliography}{29}
\raggedright


\bibitem{cherenkov1937}
P.~A.~Cherenkov,
``Visible emission of clean liquids by action of $\gamma$ radiation,''
\emph{Doklady Akademii Nauk SSSR}, vol.~2, pp.~451--454, 1934.


\bibitem{frank1937}
I.~M.~Frank and I.~E.~Tamm,
``Coherent visible radiation of fast electrons passing through matter,''
\emph{Comptes Rendus de l'Acad{\'e}mie des Sciences de l'URSS}, vol.~14, pp.~109--114, 1937.


\bibitem{jelley1958}
J.~V.~Jelley,
\emph{Cherenkov Radiation and Its Applications}.
London: Pergamon Press, 1958.


\bibitem{davut2025}
C.~Davut, G.~Xia, O.~Apsimon, J.~McGunigal, P.~Karataev, T.~Lefevre, S.~Mazzoni, and E.~Senes,
``Design and experimental verification of a bunch length monitor based on coherent Cherenkov diffraction radiation,''
\emph{Phys. Rev. Research} \textbf{7}, 013193 (2025).


\bibitem{clapp2024}
A.~Clapp, L.~Bobb, and P.~Karataev,
``Cherenkov Diffraction Radiation Beam Position Studies at Diamond Light Source,''
in \emph{Proc. 13th Int. Beam Instrum. Conf. (IBIC'24)}, Beijing, China, Sep.~2024, paper TUP33, pp.~126--129.
\doi{10.18429/JACoW-IBIC2024-TUP33}


\bibitem{benitez2024}
S.~Ben{\'i}tez, B.~Salvach{\'u}a, and M.~Chen,
``Beam loss detection based on generation of Cherenkov light in optical fibers in the CERN Linear Electron Accelerator for Research,''
\emph{Phys. Rev. Accel. Beams} \textbf{27}, 052901 (2024).
\doi{10.1103/PhysRevAccelBeams.27.052901}


\bibitem{kieffer2018}
R.~Kieffer \emph{et al.},
``Direct observation of incoherent Cherenkov diffraction radiation in the visible range,''
\emph{Phys. Rev. Lett.} \textbf{121}, 054802 (2018).
\doi{10.1103/PhysRevLett.121.054802}


\bibitem{alves2019}
D.~Alves \emph{et al.},
``Cherenkov diffraction radiation as a tool for beam diagnostics,''
in \emph{Proc. 8th Int. Beam Instrum. Conf. (IBIC'19)}, Malm{\"o}, Sweden, Sep.~2019, paper THAO01, pp.~660--664.
\doi{10.18429/JACoW-IBIC2019-THAO01}


\bibitem{mitsuhashi1999}
T.~Mitsuhashi,
``Beam profile and size measurement by SR interferometers,''
in \emph{Beam Measurement: Proceedings of the Joint US-CERN-Japan-Russia School on Particle Accelerators}, Montreux and Geneva, Switzerland, 1998, pp.~399--427.
\doi{10.1142/9789812818003_0018}


\bibitem{naito2006}
T.~Naito and T.~Mitsuhashi,
``Very small beam-size measurement by a reflective synchrotron radiation interferometer,''
\emph{Phys. Rev. ST Accel. Beams} \textbf{9}, 122802 (2006).
\doi{10.1103/PhysRevSTAB.9.122802}


\bibitem{torino2014}
L.~Torino, U.~Iriso, and T.~Mitsuhashi,
``Beam Size Measurements Using Synchrotron Radiation Interferometry at ALBA,''
in \emph{Proc. 3rd Int. Beam Instrum. Conf. (IBIC'14)}, Monterey, CA, USA, Sep.~2014, paper TUPF23, pp.~374--377.


\bibitem{vancittert1934}
P.~H.~van Cittert,
``Die Wahrscheinliche Schwingungsverteilung in Einer von Einer Lichtquelle Direkt oder Mittels Einer Linse Beleuchteten Ebene,''
\emph{Physica} \textbf{1}, 201--210 (1934).


\bibitem{bornwolf1980}
M.~Born and E.~Wolf,
\emph{Principles of Optics}, 6th ed., Chap.~10, p.~491.
Oxford: Pergamon Press, 1980.


\bibitem{karlovets2009}
D.~V.~Karlovets and A.~P.~Potylitsyn,
``Universal description for different types of polarization radiation,''
arXiv:0908.2336, 2009.
\doi{10.48550/arXiv.0908.2336}


\bibitem{malitson1965}
I.~H.~Malitson,
``Interspecimen comparison of the refractive index of fused silica,''
\emph{Journal of the Optical Society of America} \textbf{55}, 1205--1209 (1965).
\doi{10.1364/JOSA.55.001205}


\bibitem{boland2012}
M.~J.~Boland, T.~Mitsuhashi, T.~Naito, and K.~P.~Wootton,
``Intensity Imbalance Optical Interferometer Beam Size Monitor,''
in \emph{Proc. Int. Beam Instrumentation Conf. (IBIC'12)}, Tsukuba, Japan, Oct.~2012, paper WECC03, pp.~566--570.


\bibitem{novokshonov2025}
A.~Novokshonov, G.~Kube, B.~Stacey, M.~Kellermeier, T.~Vinatier, W.~Kuropka, and A.~Potylitsyn,
``Cherenkov radiation spectral and resolution properties studies at SINBAD ARES,''
in \emph{Proc. 14th Int. Beam Instrum. Conf. (IBIC'25)}, Liverpool, UK, Sep.~2025, paper TUPCO22, pp.~409--412.
\doi{10.18429/JACoW-IBIC2025-TUPCO22}

\end{thebibliography}
\end{document}